\documentclass[aps,prl,twocolumn,groupedaddress,amsmath,amssymb,showpacs]{revtex4}

\usepackage{graphicx}
\usepackage{dcolumn}
\usepackage{bm}

\begin{document}

\title{Effect of disorder on the pressure-induced superconducting state of CeAu$_{2}$Si$_{2}$}

\date{\today}
\author{Z. Ren$^{1}$}
\email{Zhi.Ren@unige.ch}
\author{G. Giriat$^{1}$}
\author{G. W. Scheerer$^{1}$}
\author{G. Lapertot$^{2}$}
\author{D. Jaccard$^{1}$}
\affiliation{$^{1}$DQMP - University of Geneva, 24 Quai Ernest-Ansermet, 1211 Geneva 4, Switzerland}
\affiliation{$^{2}$SPSMS, UMR-E CEA/UJF-Grenoble 1, INAC, Grenoble, F-38054, France}

\begin{abstract}
 CeAu$_{2}$Si$_{2}$ is a newly discovered pressure-induced heavy fermion superconductor which shows very unusual interplay between superconductivity and magnetism under pressure. Here we compare the results of high-pressure measurements on single crystalline CeAu$_{2}$Si$_{2}$ samples with different levels of disorder. It is found that while the magnetic properties are essentially sample independent, superconductivity is rapidly suppressed when the residual resistivity of the sample increases.
 We show that the depression of bulk $T_{\rm c}$ can be well understood in terms of pair breaking by nonmagnetic disorder,
 which strongly suggests an unconventional pairing state in pressurized CeAu$_{2}$Si$_{2}$. Furthermore, increasing the level of disorder leads to the emergence of another phase transition at $T^{*}$ within the magnetic phase, which might be in competition with superconductivity.
\end{abstract}
\pacs{74.62.Fj, 74.62.En, 74.70.Tx}

\maketitle
\maketitle
\section{I. Introduction}
Ce-based magnetic compounds that become superconducting under pressure have attracted a lot of attention because of the intimate connection between superconductivity (SC) and magnetic or valence instabilities \cite{ReviewKnebel}.
Prominent examples include CeCu$_{2}$Ge$_{2}$ \cite{CeCu2Ge2Jaccard}, CePd$_{2}$Si$_{2}$ \cite{CePd2Si2}, CeIn$_{3}$ \cite{CeIn3} and CeRhIn$_{5}$ \cite{CeRhIn5}.
Very recently, pressure-induced heavy fermion SC with transition temperatures $T_{\rm c}$ up to 2.5 K is observed in the antiferromagnet CeAu$_{2}$Si$_{2}$ \cite{CeAu2Si2}, which is both isostructural and isoelectronic to the first unconventional superconductor CeCu$_{2}$Si$_{2}$ \cite{CeCu2Si2}. It is quite remarkable in CeAu$_{2}$Si$_{2}$ that SC coexists with long-range magnetic order over a huge pressure interval of 11 GPa. Moreover, in approximately one-third of this pressure range, the magnetic ordering temperature $T_{\rm M}$ and $T_{\rm c}$ are simultaneously enhanced by pressure \cite{CeAu2Si2}. These behaviors are hardly explained within the common scenarios of Cooper pairing mediated by spin \cite{spinmediatedpairing} or valence-fluctuations \cite{valencemediatedpairng}, and thus it is of particular interest to clarify the nature of SC in this material.

The $T_{\rm c}$ response to the level of nonmagnetic disorder is known to provide useful information for the phase of the superconducting gap function. For conventional $s$-wave superconductors, no pair breaking is expected by nonmagnetic disorder as long as the system remains metallic, according to the Anderson's theorem \cite{Anderson}. By contrast, for non $s$-wave superconductors, in which there is a sign reversal in the superconducting gap function, scattering from nonmagnetic disorder averages out the gap over the Fermi surface and results in a strong suppression of $T_{\rm c}$. This effect has been observed in a number of unconventional superconductors, such as UPt$_{3}$ \cite{UPt3}, YBaCu$_{3}$O$_{6+x}$ \cite{cuprate}, Sr$_{2}$RuO$_{4}$ \cite{Sr2RuO4}, BEDT-TTF salts \cite{organic} and CePt$_{3}$Si \cite{CePt3Si}. Although it is commonly believed that the pairing symmetry is non $s$-wave for Ce-based pressure-induced superconductors \cite{Lonzarich}, there is little systematic study of a similar effect at high pressure.

In this paper, we present the pressure responses of two CeAu$_{2}$Si$_{2}$ crystals grown from different fluxes with in-plane residual resistivities $\rho_{\rm 0}$ = 1.8 and 12.2 $\mu$$\Omega$cm, respectively.
The results show that while the critical pressures for the disappearance of magnetism and the delocalization of the Ce 4$f$ electrons are almost independent on $\rho_{\rm 0}$, the high-$\rho_{\rm 0}$ sample shows a much narrower pressure range for SC and a considerably lower maximum $T_{\rm c}$.
A detailed analysis indicates that at $\sim$21.2 GPa, SC with an initial onset $T_{\rm c}$ of $\sim$2.5 K is destroyed when $\rho_{\rm 0}$ exceed $\sim$46 $\mu\Omega$cm, i.e. when the carrier mean free path is reduced to be similar to the superconducting coherence length.
Since there is good evidence that $\rho_{\rm 0}$ is dominated by the contribution of nonmagnetic disorder, our results point to unconventional SC in CeAu$_{2}$Si$_{2}$ under pressure. In addition, the high-$\rho_{\rm 0}$ sample displays another phase transition at a temperature below $T_{\rm M}$, which is probably competing with SC.

\section{II. Experimental}
Crystal growth of CeAu$_{2}$Si$_{2}$ samples by Sn flux and Au-Si self-flux are described in detail in Ref. \cite{CeAu2Si2}. The resulting crystals are labeled hereafter as CeAu$_{2}$Si$_{2}$(Sn) and CeAu$_{2}$Si$_{2}$(self), respectively. Within the resolution limits of x-ray and microprobe techniques, no difference is observed in the crystal structure and chemical composition of the Sn- and self-flux samples.

High pressure experiments were performed using a Bridgman-type sintered diamond-anvil cell with steatite as soft-solid pressure medium and lead (Pb) as pressure gauge \cite{Holmesiron}.
The results of high-pressure experiments on CeAu$_{2}$Si$_{2}$(Sn) have been reported in Ref. \cite{CeAu2Si2}.
For CeAu$_{2}$Si$_{2}$(self), measurements are carried out in two different pressure cells.
In the first pressure cell, only resistivity is measured up to 25.5 GPa.
The second pressure cell is designed to measure both resistivity and ac heat capacity, but the pressure is limited to 20.5 GPa.
In both cells, the CeAu$_{2}$Si$_{2}$(self) sample with its $ab$ plane perpendicular to the compressive force is connected in series with the Pb gauge.
The resistivity was measured by using a standard four-probe method.
For ac-calorimetry measurements, a chromel wire, which is otherwise used as a voltage lead, serves as the heater, and the sample temperature oscillations are detected by a Au/AuFe(0.07\%) thermocouple. More details of the measurement procedure and data analysis can be found in Ref. \cite{Holmesiron}. A good agreement is found between the results of the two cells, indicating that they reflect the intrinsic properties of the sample.
\section{III. Results and Discussion}
\subsection{A. Ambient pressure results}
\begin{figure}
\includegraphics*[width=7.5cm]{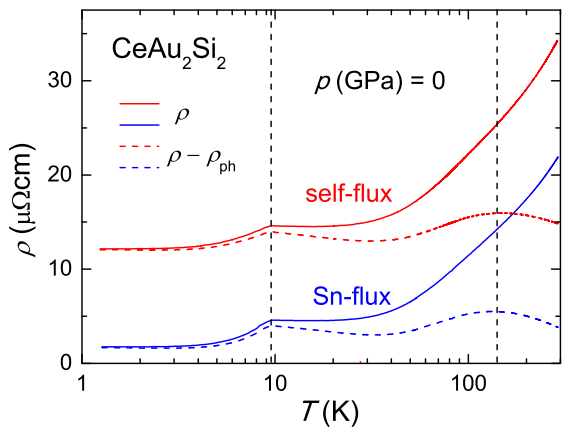}
\caption{(Color online)
Logarithmic temperature dependence of the in-plane resistivity at ambient pressure before (solid lines) and after (dashed lines) subtraction of the phonon contribution for the CeAu$_{2}$Si$_{2}$ crystals grown from the different fluxes.
The vertical dotted lines are a guide to the eyes.
}
\label{fig1}
\end{figure}

Figure 1 shows the comparison of the ambient pressure resistivity data of the CeAu$_{2}$Si$_{2}$ crystals grown from the different fluxes.
It can be observed that the resistivity curve of CeAu$_{2}$Si$_{2}$(self) is an almost rigid upshift of that of CeAu$_{2}$Si$_{2}$(Sn).
After subtraction of the phonon contribution to the resistivity ($\rho_{\rm ph}$), which is assumed to be linear in temperature and pressure independent,
both samples exhibit a resistivity maximum at $\sim$140 K and a sharp drop in resistivity due to the magnetic ordering below $T_{N}$ $\approx$ 10 K. Furthermore, $\rho_{\rm ph}$ $\approx$ 0.067$T$ ($\mu$$\Omega$cm) estimated for CeAu$_{2}$Si$_{2}$(self) is in agreement with that of CeAu$_{2}$Si$_{2}$(Sn) \cite{CeAu2Si2} within the geometrical factor uncertainty ($\sim$ 10\%).
Hence the resistivities of the two crystals differ only by their $\rho_{\rm 0}$ values, which correspond to different degrees of disorder.

\subsection{B. Pressure response of CeAu$_{2}$Si$_{2}$(self)}
\begin{figure}
\includegraphics*[width=8cm]{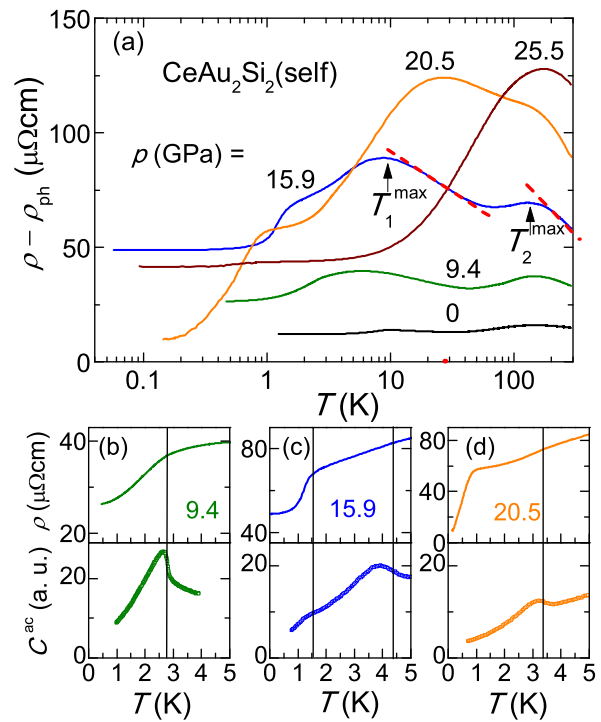}
\caption{(Color online)
 (a) For typical pressures, logarithmic temperature dependence of in-plane resistivity of the self-flux grown CeAu$_{2}$Si$_{2}$(self) crystals after subtraction of the phonon contribution. The two characteristic maxima $T_{\rm 1}^{\rm max}$ and $T_{\rm 2}^{\rm max}$ at 15.9 GPa are marked by arrows.
 The dashed lines are a guide to the eyes, evidencing the $-$ln$T$ behavior.
 Panels (b)$-$(d) show the comparison of the resistivity and ac heat capacity for three different pressures.
 The solid lines are a guide to the eyes.
}
\label{fig2}
\end{figure}

\begin{figure}
\includegraphics*[width=7.5cm]{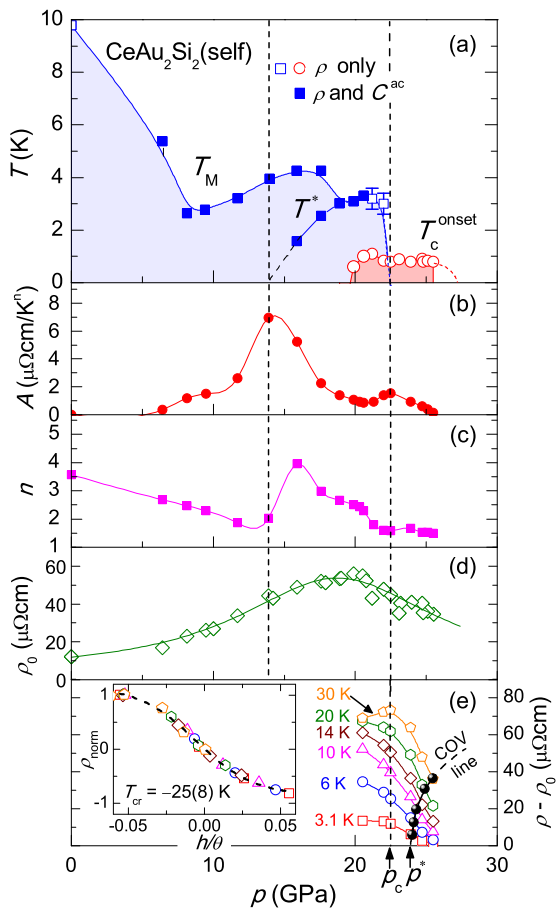}
\caption{(Color online)
(a) Experimental $p$-$T$ phase diagram of CeAu$_{2}$Si$_{2}$(self). $T_{\rm c}^{\rm onset}$ and $T_{\rm M}$ represent the superconducting transition onset and the magnetic ordering temperatures, respectively.
The open (closed) symbols denote the data extracted from the resistivity measurements only (both resistivity and heat capacity measurements).
(b), (c) and (d) show the pressure dependencies of the coefficient $A$, temperature exponent $n$, and residual resistivity $\rho_{\rm 0}$, respectively, obtained from the fitting of the power law $\rho(T)$ = $\rho_{\rm 0}$ + $AT^{n}$ to the low temperature resistivity data.
(e) Plot of $\rho^{*}$ = $\rho$ -- $\rho_{0}$ versus $p$ at selected temperatures up to 30 K. The closed circles indicate for each isotherm the 50\% drop of $\rho^{*}$ compared to its value at 22.5 GPa and define the crossover (COV) line.
The inset shows the collapse of all normalized data $\rho_{\rm norm}$ when plotted as a function of the generalized distance $h/\theta$ from the critical end point located at $p^{*}$ $\approx$ 23.9 GPa and $T_{\rm cr}$ = $-$25 K.
The vertical dashed lines are a guide to the eyes.
The two critical pressures $p_{\rm c}$ and $p^{*}$ are indicated by labeled arrows.
}
\label{fig3}
\end{figure}
Typical results at selected pressures of the resistivity ($\rho$) and ac heat capacity ($C^{\rm ac}$) of CeAu$_{2}$Si$_{2}$(self) are shown in Fig. 2.
Apart from a much larger residual resistivity $\rho_{\rm 0}$, the overall behavior of the non-phononic resistivity [Fig. 2(a)] is very similar to that of CeAu$_{2}$Si$_{2}$(Sn). At the intermediate pressure of 15.9 GPa, two broad maxima exist at $T_{\rm 1}^{\rm max}$ and $T_{\rm 2}^{\rm max}$, and above each maxima the data follow a $-$ln$T$ dependence, which manifests the incoherent Kondo scattering of the ground state and excited crystal-filed (CF) levels, respectively \cite{scattering1,twomaxima}. As pressure is increased to 20.5 GPa, $T_{\rm 1}^{\rm max}$ almost doubles while $T_{\rm 2}^{\rm max}$ remains nearly unchanged. At the highest pressure of 25.5 GPa, the Kondo effect dominates over the CF splitting so that the two maxima are already merged into a single peak at $\sim$180 K.
Concomitantly, both the magnitude of the resistivity and the $-$ln$T$ slope increase rapidly with pressure, signifying a strong enhancement of the Kondo interaction under pressure.

In Fig. 2(b)$-$(d), we compare the results of $\rho$ and $C^{\rm ac}$ below 5 K at three typical pressures.
At 9.4 GPa, the change of slope in resistivity at $\sim$2.7 K coincides with the midpoint of the sharp jump in $C^{\rm ac}$($T$), indicating a magnetic ordering \cite{determinationofTM}.
Notably, at 15.9 GPa two jumps in $C^{\rm ac}$($T$) are observed. The one at $\sim$ 4.4 K corresponds to a slight slope change of the resistivity,
while the other at $\sim$ 1.5 K is accompanied by a steep resistivity drop that is independent of the applied current and is not due to SC.
Thus it appears that at this pressure the sample undergoes two successive phase transitions, similarly to CeCu$_{2}$Ge$_{2}$  at $\sim$3 GPa \cite{CeCu2Ge2-1}.
At 20.5 GPa, the highest pressure at which $C^{\rm ac}$ is measured, only one magnetic transition is detectable in both $\rho$ and $C^{\rm ac}$($T$) at 3.3 K, and, below 1 K the incomplete resistive transition without any corresponding anomaly in $C^{\rm ac}$($T$) indicates SC of filamentary nature.

The resulting pressure-temperature phase diagram is shown in Fig. 3(a).
The magnetic ordering temperature $T_{\rm M}$ initially decreases with increasing pressure, as expected, due to the enhancement of the Kondo interaction.
However, $T_{\rm M}$ already starts to increase with pressure above 8 GPa. At 15.9 GPa, another transition appears at $T^{*}$ $<$ $T_{\rm M}$.
With further increasing pressure, $T^{*}$ rises while $T_{\rm M}$ shows a maximum, and the two transitions merge at 18.9 GPa. At higher pressures, $T_{\rm M}$ exhibits a dome-shaped dependence and finally disappears abruptly above 22.5 GPa. On the other hand, SC is observed from 19.9 GPa up to the highest investigated pressure. It is pointed out that zero resistivity is achieved only at 21.2 GPa, the pressure at which the onset $T_{\rm c}$ reaches its maximum of $\sim$1.1 K. Both $T_{\rm c}$ and $T_{\rm M}$ are enhanced within a narrow pressure range between 19.9 and 20.6 GPa.

Figure 3(b)-(d) shows the fitting parameters of the power law $\rho(T)$ = $\rho_{\rm 0}$ +$AT^{n}$ to the resistivity data plotted as a function of pressure. Thanks to a sufficiently broad temperature window between $T_{\rm c}$ and $T_{\rm M}$, we are able to extract reliable parameters for the whole pressure range.
The $A$ coefficient exhibits two maxima at 13.9 and 22.5 GPa, respectively. The latter together with a minimum $n$ exponent ($n$ $\approx$ 1.5) coincides with the disappearance of the magnetic order, indicating a magnetic quantum critical point at $p_{\rm c}$ = 22.5 $\pm$ 0.5 GPa. However, the former with $n$ $\approx$ 2 occurs at a pressure close to that of the interpolation of $T^{*}$ to 0 K, which points to the possibility of a putative quantum phase transition occurring within the magnetic phase. Actually, $\rho_{\rm 0}$ shows a broad peak at $\sim$20 GPa, suggesting that the maximum scattering rate happens in between these two QCPs. Nevertheless, the large $A$ value at 13.9 GPa may contain a significant contribution from the electron-magnon scattering, and thus provides little information of the effective mass.

Figure 3(e) shows the plot of isothermal resistivity $\rho^{*}(p)$ = $\rho(p)$ $-$ $\rho_{\rm 0}(p)$ versus $p$ at selected temperatures up to 30 K.
Above 22.5 GPa, $\rho^{*}$($p$) decreases steeply with pressure, revealing the continuous delocalization of the Ce 4$f$ electrons.
For the data analyses, we follow the procedure described in Ref. \cite{CeCu2Si2seyfarth}, which is based on the assumption of an underlying critical end point located at ($p_{\rm cr}$, $T_{\rm cr}$) in the $p$-$T$ plane.
It turns out that all the normalized resistivity curves $\rho_{\rm norm}$($p$) = ($\rho^{*}$($p$) $-$ $\rho^{*}$($p_{\rm 50\%}$))/$\rho^{*}$($p_{\rm 50\%}$) below 30 K fall on a single curve when plotted against $h/\theta$,
where for each temperature, $p_{\rm 50\%}$ denotes the pressure corresponding to the midpoint of the $\rho^{*}$($p$)-drop compared to its value at 22.5 GPa, $h$ = ($p$ $-$ $p_{\rm 50\%}$)/$p_{\rm 50\%}$ and $\theta$ = ($T$ $-$ $T_{\rm cr}$)/$|T_{\rm cr}|$ with the only free parameter $T_{\rm cr}$ = $-$25(8) K. The negative $T_{\rm cr}$ indicates a crossover rather than a first-order transition. Moreover, the extrapolation of the temperature dependence of $p_{\rm 50\%}$ to 0 K yields the critical end pressure $p^{*}$($\approx$ $p_{\rm cr}$) = 23.9 $\pm$ 0.7 GPa.
\begin{figure}
\includegraphics*[width=6.5cm]{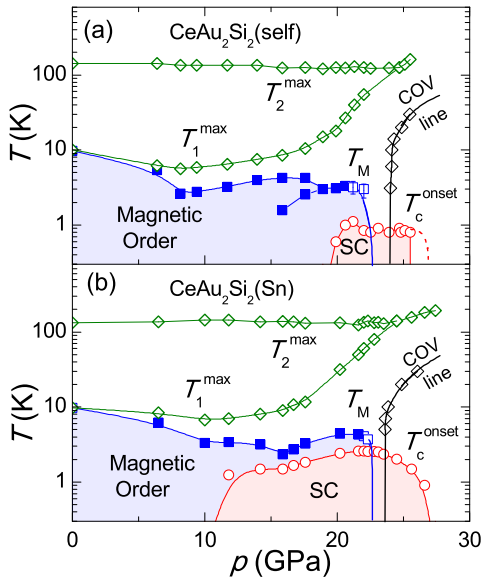}
\caption{(Color online)
$p-T$ phase diagrams of (a) CeAu$_{2}$Si$_{2}$(self) and (b) CeAu$_{2}$Si$_{2}$(Sn) including the temperatures $T_{\rm 1}^{\rm max}$ and $T_{\rm 2}^{\rm max}$.
Note that the vertical axis is in logarithmic scale. The crossover(COV) line is determined by the scaling analysis of the resistivity.
}
\label{fig4}
\end{figure}

\subsection{C. Comparison with the results of CeAu$_{2}$Si$_{2}$(Sn)}
Figure 4 shows the $p$-$T$ phase diagrams of CeAu$_{2}$Si$_{2}$(self) and CeAu$_{2}$Si$_{2}$(Sn) including the lines defined by the temperatures $T_{\rm 1}^{\rm max}$ and $T_{\rm 2}^{\rm max}$ of the resistivity maxima in the paramagnetic phase, as well as the crossover (COV) lines obtained by the 50\% drop of $\rho^{*}$. Clearly, these lines are almost identical for both crystals.
Since the temperatures $T_{\rm 1}^{\rm max}$ and $T_{\rm 2}^{\rm max}$ (for $p$ $>$ 15.9 GPa) scale approximately with the Kondo temperature and CF splitting energy respectively \cite{twomaxima}, it is obvious that the pressure evolution of the characteristic high-temperature energy scales are essentially sample independent.
This reflects that both the Kondo interaction and CF levels depend mainly on the local environment of the Ce ions.

By contrast, at temperatures below 5 K, the two phase diagrams show significant differences.
Although the critical pressures $p_{\rm c}$ and $p^{*}$ are nearly identical for both samples, in CeAu$_{2}$Si$_{2}$(self) a much higher pressure (19.9 GPa) is needed to induce SC and the maximum onset $T_{\rm c}$ is reduced by a factor of 2.3.
As a consequence, the pressure range for the overlap between the magnetic and superconducting phases is restricted to $\sim$3 GPa.
It is also noteworthy that the pressure evolution of the magnetic order is more complex for CeAu$_{2}$Si$_{2}$(self).
While the origin of the transition at $T^{*}$ remains unclear, it is possible that the resulting ground state is competing with SC, as will be discussed further below.

\begin{figure}
\includegraphics*[width=6.3cm]{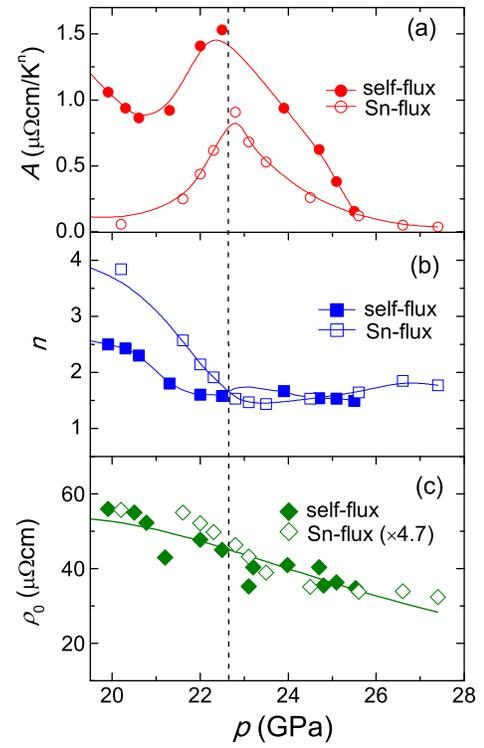}
\caption{(Color online)
Comparison of the pressure dependencies of the fitting parameters, (a) the $A$ coefficient, (b) the $n$ exponent, (c) the residual resistivity $\rho_{\rm 0}$ for CeAu$_{2}$Si$_{2}$(self) and CeAu$_{2}$Si$_{2}$(Sn).
The open and closed symbols denote the data for CeAu$_{2}$Si$_{2}$(Sn) and CeAu$_{2}$Si$_{2}$(self), respectively.
Note that the $\rho_{\rm 0}$ data for CeAu$_{2}$Si$_{2}$(Sn) are multiplied by a factor of 4.7.
The vertical dashed line is a guide to the eyes.
}
\label{fig5}
\end{figure}

Figure 5 compares the power-law fitting parameters of the resistivity data above $T_{\rm c}$ for CeAu$_{2}$Si$_{2}$(self) and CeAu$_{2}$Si$_{2}$(Sn).
It can be noted that the pressure dependencies of the three parameters are very similar in both cases.
The maximum A coefficient and minimum $n$($<$ 2) exponent at $p_{\rm c}$ are typical for a QCP.
Above $p_{\rm c}$, while the $n$ values increase only slightly, the $A$ values drop abruptly by more than one order of magnitude, indicating a transition from a strongly to a weakly correlated regime.
Moreover, the $\rho_{\rm 0}$ data of CeAu$_{2}$Si$_{2}$(Sn), when multiplied by a factor of 4.7, match well with those of CeAu$_{2}$Si$_{2}$(self).
This scaling factor is not far from its ambient pressure value ($\sim$6.7), suggesting that the difference in the levels of disorder between CeAu$_{2}$Si$_{2}$(self) and CeAu$_{2}$Si$_{2}$(Sn) does not change much with pressure.

\begin{figure}
\includegraphics*[width=6.2cm]{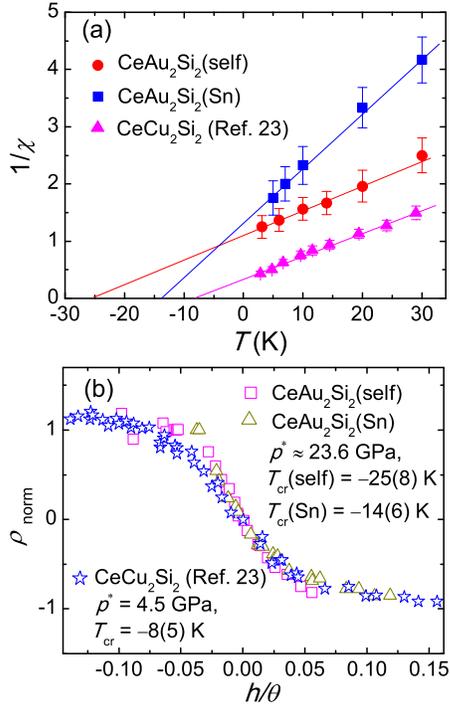}
\caption{(Color online)
(a) Temperature dependencies of the inverse slope 1/$\chi$ (see text for details) for CeAu$_{2}$Si$_{2}$(self), CeAu$_{2}$Si$_{2}$(Sn), and CeCu$_{2}$Si$_{2}$ from Ref. \cite{CeCu2Si2seyfarth}.
$T_{\rm cr}$ values are extracted from linear fits to the data (solid lines).
(b) $\rho_{\rm norm}$ data plotted as a function of $h/\theta$ for the three cases, showing a good agreement.
}
\label{fig6}
\end{figure}
We next turn our attention to the scaling behavior of the resistivity near $p^{*}$.
Figure 6(a) shows the temperature dependence of 1/$\chi$, where $\chi$ = $|$$d\rho_{\rm norm}$/$dp$$|$$_{p_{\rm 50\%}}$ is the slope of the resistivity drop at the midpoint, for CeAu$_{2}$Si$_{2}$(self) and CeAu$_{2}$Si$_{2}$(Sn) \cite{CeAu2Si2}, as well as CeCu$_{2}$Si$_{2}$ \cite{CeCu2Si2seyfarth} for comparison. In the three cases, 1/$\chi$ diminishes on lowering temperature, indicating that the slope becomes increasingly steep as the systems approach their critical end point located at ($p_{\rm cr}$, $T_{\rm cr}$).
Assuming 1/$\chi$ $\propto$ ($T$ $-$ $T_{\rm cr}$), $T_{\rm cr}$ can be obtained by a linear extrapolation of the data to the $x$-axis.
Remarkably, despite the large differences in $T_{\rm cr}$ ranging from $-$25 K to $-$8 K, the $\rho_{\rm norm}$ data below 30 K follow almost the same curve when plotted as a function of $h/\theta$ especially for $h/\theta$ $>$ 0 ($p$ $>$ $p_{\rm 50\%}$), as shown in Fig. 6(b). This means that,
for a generalized distance $h/\theta$ from the critical end point, $\rho_{\rm norm}$ behaves in the same way for both CeAu$_{2}$Si$_{2}$ and CeCu$_{2}$Si$_{2}$.
Further study is needed to establish the level of generality of such a behavior in related Ce-based compounds.
For our two CeAu$_{2}$Si$_{2}$ crystals, it appears that higher (less negative) $T_{\rm cr}$ corresponds to higher $T_{\rm c}$.
However, despite their similar $T_{\rm c}$ values near their respective $p^{*}$, the absolute $T_{\rm cr}$ value of CeAu$_{2}$Si$_{2}$(Sn) is nearly twice that of CeCu$_{2}$Si$_{2}$.
Nevertheless, the $T_{\rm cr}$ value of CeAu$_{2}$Si$_{2}$ could be considerably underestimated due to the unavoidable degradation of hydrostatic conditions at very high pressure.

\begin{figure}
\includegraphics*[width=6.5cm]{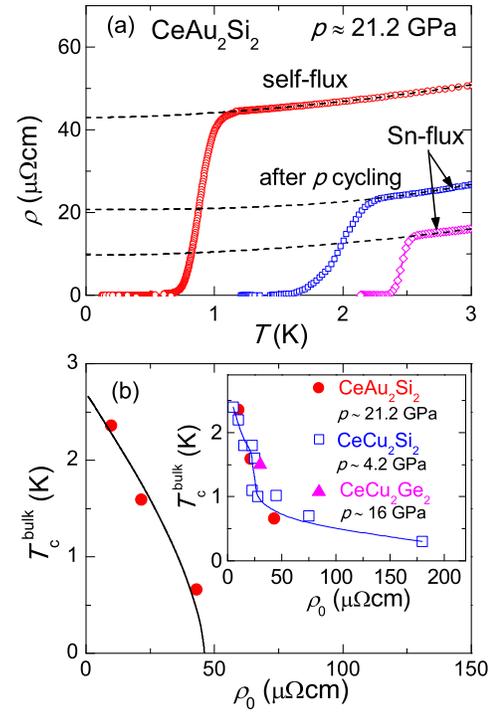}
\caption{(Color online)
(a) Temperature dependencies of the resistivity of CeAu$_{2}$Si$_{2}$ at $\sim$21.2 GPa for different level of disorder. The dashed lines denote the power-law fit employed to extract the residual resistivity $\rho_{\rm 0}$.
(b) Plot of $T_{\rm c}^{\rm bulk}$ versus $\rho_{\rm 0}$ for CeAu$_{2}$Si$_{2}$ at $\sim$21.2 GPa. The solid line is a fit from the AG theory to the data.
The inset shows $T_{\rm c}^{\rm bulk}$ plotted as a function of $\rho_{\rm 0}$ for CeAu$_{2}$Si$_{2}$ at $\sim$21.2 GPa, CeCu$_{2}$Si$_{2}$ at $\sim$4.2 GPa and CeCu$_{2}$Ge$_{2}$ at $\sim$16 GPa.
Note that we have collected $all$ available data from Refs. \cite{valencemediatedpairng,Holmesthesis,CeCu2Ge2-1,CeCu2Si2seyfarth,Holmesreview,BellarbiCeCu2Si2,Emmathesis}.
}
\label{fig7}
\end{figure}
To gain insight into the pressure-induced SC in CeAu$_{2}$Si$_{2}$, we show in Fig. 7(a) the resistivity below 3 K of the Sn- and self-flux grown crystals around 21.2 GPa, a pressure close to that of the optimum $T_{c}$.
As can be seen, CeAu$_{2}$Si$_{2}$(Sn) has a low $\rho_{\rm 0}$ and shows a sharp superconducting transition below 2.55 K with a width of 0.18 K.
By contrast, CeAu$_{2}$Si$_{2}$(self) has a much higher $\rho_{\rm 0}$, and its superconducting onset temperature is shifted to 1.15 K while the transition width increases to 0.55 K.
This trend is further corroborated by investigating the effect of pressure cycling on CeAu$_{2}$Si$_{2}$(Sn), which tends to induce further disorder(dislocation) and therefore increase $\rho_{\rm 0}$. Indeed, when pressure is increased again up to $\sim$21.2 GPa (after a partial depressurization from 27.6 down to 10 GPa),
the $\rho_{\rm 0}$ value doubles while the onset $T_{\rm c}$ decreases to $\sim$2.2 K. Concomitantly, the resistive transition becomes much broader most likely due to the decrease of pressure homogeneity, which is inevitable after the pressure cycling.

In Fig. 7(b), we plot the dependence of $T_{\rm c}^{\rm bulk}$ on $\rho_{\rm 0}$ at $\sim$21.2 GPa.
Here $T_{\rm c}^{\rm bulk}$ is defined as the temperature where zero resistivity is achieved,
given that the completeness of resistive transition coincides with the jump in the ac heat capacity \cite{CeAu2Si2}.
As can be seen,  $T_{\rm c}^{\rm bulk}$ decreases rapidly with increasing $\rho_{\rm 0}$.
For $f$-electron systems, $\rho_{\rm 0}$ can be expressed as
$\rho_{\rm 0}$ = $\rho_{\rm 0}^{\rm Born}$ + $\rho_{\rm 0}^{\rm unit}$, where $\rho_{\rm 0}^{\rm Born}$ and $\rho_{\rm 0}^{\rm unit}$ are due to the nonmagnetic disorder of non-$f$ elements and the defects of Ce ions, respectively \cite{Holmesreview}. Under pressure, $\rho_{\rm 0}^{\rm unit}$ remains essentially unaffected while $\rho_{\rm 0}^{\rm Born}$ is subject to a large enhancement due to critical fluctuations \cite{Miyake}.
The values of $\rho_{\rm 0}$ for both CeAu$_{2}$Si$_{2}$(self) and CeAu$_{2}$Si$_{2}$(Sn) are much larger at $\sim$21.2 GPa than at ambient pressure, indicating that $\rho_{\rm 0}^{\rm Born}$ dominates $\rho_{\rm 0}$.
Therefore, our results are consistent with pair breaking by nonmagnetic disorder.

According to the Abrikosov-Gor'kov (AG) theory \cite{AG} generalized for nonmagnetic disorder in a non-$s$-wave superconductor \cite{AG1,AG2,AG3}, the suppression of $T_{\rm c}$ follows ln($\frac{T_{\rm c0}}{T_{\rm c}}$) =  $\Psi(\frac{1}{2} + \frac{\alpha T_{\rm c0}}{T_{\rm c}})$ $-$ $\Psi(\frac{1}{2})$, where $\Psi$ is the digamma function, $\alpha$ = $\hbar$/(2$\pi$$k_{\rm B}$$\tau$$T_{\rm c0}$), $\tau$ is the scattering time, and $T_{\rm c0}$ is $T_{\rm c}$ in the disorder-free limit ($\alpha$ $\rightarrow$ 0). The model predicts that $T_{\rm c}$ vanishes at a critical $\alpha$, which is roughly equivalent to the fact that the carrier mean free path $l$ is comparable to the superconducting (Pippard) coherence length $\xi_{\rm 0}$. It turns out that the experimental data obey well the functional form of the AG theory, with $T_{\rm c0}^{\rm bulk}$ $\approx$ 2.7 K and a critical $\rho_{\rm 0}^{\rm cr}$ $\sim$ 46 $\mu$$\Omega$cm.
The corresponding critical mean free path $l^{\rm cr}$ can be estimated using the relation $l$ = (1.27 $\times$ 10$^{4}$)/$\rho_{\rm 0}$($N / V$)$^{2/3}$, where $\rho_{\rm 0}$ is in $\Omega$cm, $N$ is the number of conduction electrons per unit cell, and $V$ is the unit cell volume in cm$^{3}$ \cite{UBe13}. Assuming $N$ = 6 (there are two formula units per unit cell) and
with $V$ $\approx$ 1.6 $\times$ 10$^{-22}$ cm$^{3}$ \cite{CeAu2Si2}, we obtain $l^{\rm cr}$ $\approx$ 27 {\AA}, which is half of the Ginzburg-Landau(GL) coherence length $\xi_{\rm GL}$(0) = 55 {\AA} ($\approx$ $\xi_{\rm 0}$) deduced from the measurement of the upper critical field at 22.3 GPa with $T_{\rm c}$ $\sim$ 2.5 K \cite{CeAu2Si2}. It should be pointed out that $l^{\rm cr}$ could be underestimated due to the following reasons: (i) The $N$ value is overestimated; (ii) Only the parts of the sample with a low enough $\rho_{\rm 0}$ become superconducting such that $\rho_{\rm 0}^{\rm cr}$ is overestimated; (iii) The calculated $l^{\rm cr}$ reflects mainly the contribution from the scattering of the light quasiparticles, while $l^{\rm cr}$ for the heavy quasiparticles that form Cooper pairs could be longer \cite{CeCuSiGedisorder}. Thus it is reasonable to speculate that the actual $l^{\rm cr}$ is close to $\xi_{\rm 0}$.

The above analysis shows that the suppression of bulk $T_{\rm c}$ can be understood, not only qualitatively but also quantitatively, within the pair breaking model.
This strongly points to unconventional pairing in the superconducting state of CeAu$_{2}$Si$_{2}$ under pressure. As a matter of fact, in CeCu$_{2}$Si$_{2}$ at a similar volume $V$ ($p$ $\sim$ 4.2 GPa) \cite{CeAu2Si2}, the $\rho_{\rm 0}$ dependence of the maximum $T_{\rm c}^{\rm bulk}$ shows very similar behavior to that of CeAu$_{2}$Si$_{2}$ for $\rho_{\rm 0}$ $<$ 50 $\mu\Omega$cm [inset of Fig. 7(b)] \cite{valencemediatedpairng,Holmesthesis,CeCu2Ge2-1,CeCu2Si2seyfarth,Holmesreview,BellarbiCeCu2Si2,Emmathesis}, suggesting a common mechanism of SC in these compounds. Moreover, at this $V$ magnetic order is still present in CeAu$_{2}$Si$_{2}$ but is absent in CeCu$_{2}$Si$_{2}$, which supports the idea that magnetic order and SC are not directly related \cite{CeAu2Si2}. For $\rho_{\rm 0}$ $>$ 50 $\mu\Omega$cm, the depression of $T_{\rm c}^{\rm bulk}$ of CeCu$_{2}$Si$_{2}$ with increasing $\rho_{\rm 0}$ becomes much weaker, and SC survives even when $\rho_{\rm 0}$ is of the order of the Ioffe-Regel limit \cite{CeCu2Si2disorder}. However, the large $\rho_{\rm 0}$ could be due to the effect of Kondo holes \cite{Kondohole1,Kondohole2}, and thus is no longer a good indication of the level of disorder.

The sensitivity of SC in CeAu$_{2}$Si$_{2}$ to nonmagnetic disorder also allows of an explanation for the transition at $T^{*}$ observed in CeAu$_{2}$Si$_{2}$(self).
It is theoretically demonstrated that when coupled to quantum critical fluctuations, disorder can induce regions of local order or even long range order in the host phase \cite{localorder}.
Indeed, a recent experimental study shows that nonmagnetic Cd impurities in CeCoIn$_{5}$ nucleate magnetic regions even when global magnetic order is suppressed and bulk SC is restored by pressure, which is ascribed to the local competition between magnetism and SC \cite{CdinCeCoIn5}.
As shown above, the emergence of the transition at $T^{*}$ may be related to a putative QCP, and almost coincides with the establishment of bulk SC in CeAu$_{2}$Si$_{2}$(Sn).
It is possible that near the disorder sites in CeAu$_{2}$Si$_{2}$(Sn), SC is locally destroyed and regions of competing order are formed. With increasing disorder, these regions are expected to grow in size and start to overlap.
As a result, above a certain level of disorder, SC is destroyed completely and long range order develops, as in CeAu$_{2}$Si$_{2}$(self).

\section{IV. Conclusion}
In summary, we have studied the effect of disorder on the pressure-induced heavy fermion superconductor CeAu$_{2}$Si$_{2}$ through the comparison of high-pressure results from single crystalline samples with two different $\rho_{\rm 0}$ values.
It is found that, with the increase of $\rho_{\rm 0}$, both the pressure range for SC and the maximum $T_{\rm c}$ are reduced, although the critical behaviors near the magnetic-nonmagnetic boundary and the delocalization of Ce 4$f$ electrons under pressure are essentially unaffected. The bulk $T_{\rm c}$ dependence on $\rho_{\rm 0}$ near the optimum pressure for SC is very similar to that of CeCu$_{2}$Si$_{2}$, and is consistent with the pair breaking effect by nonmagnetic disorder. These results not only provide evidence for unconventional SC in CeAu$_{2}$Si$_{2}$ under pressure, but also suggest that the two Ce$X$$_{2}$Si$_{2}$ ($X$ = Cu or Au) compounds share a common pairing mechanism. Finally, for the sample with a higher $\rho_{\rm 0}$ value, a new phase transition appears at $T^{*}$ below $T_{\rm M}$, which is probably related to an order that competes with SC.
In this respect, the clarification of the nature of the transition at $T^{*}$, which may help to elucidate the pairing mechanism, is worth pursuing in future studies.

\section{Acknowledgement}
We acknowledge technical assistance from M. Lopes, and financial support from the Swiss National Science Foundation through Grant No. 200020-137519.


\begin{thebibliography}{99}
\expandafter\ifx\csname natexlab\endcsname\relax\def\natexlab#1{#1}\fi
\expandafter\ifx\csname bibnamefont\endcsname\relax
  \def\bibnamefont#1{#1}\fi
\expandafter\ifx\csname bibfnamefont\endcsname\relax
  \def\bibfnamefont#1{#1}\fi
\expandafter\ifx\csname citenamefont\endcsname\relax
  \def\citenamefont#1{#1}\fi
\expandafter\ifx\csname url\endcsname\relax
  \def\url#1{\texttt{#1}}\fi
\expandafter\ifx\csname urlprefix\endcsname\relax\def\urlprefix{URL }\fi
\providecommand{\bibinfo}[2]{#2}
\providecommand{\eprint}[2][]{\url{#2}}

\bibitem{ReviewKnebel}
G. Knebel, D. Aoki, and J. Flouquet,
Comptes Rendus Physique {\bf 12}, 542 (2011).

\bibitem{CeCu2Ge2Jaccard}
D. Jaccard, K. Behnia, and J. Sierro,
Phys. Lett. A {\bf 163}, 475 (1992).

\bibitem{CePd2Si2}
F. M. Grosche, S. R. Julian, N. D. Mathur, and G. G. Lonzarich,
Physica {\bf 223-224B}, 50 (1996).

\bibitem{CeIn3}
N. D. Mathur, F. M. Grosche, S. R. Julian, I. R. Walker, D. M. Freye, R. K. W. Haselwimmer, and G. G. Lonzarich,
Nature (London) {\bf 394}, 39 (1998).

\bibitem{CeRhIn5}
H. Hegger, C. Petrovic, E. G. Moshopoulou, M. F. Hundley, J. L. Sarrao, Z. Fisk, and J. D. Thompson,
Phys. Rev. Lett. {\bf 84}, 4986 (2000).

\bibitem{CeAu2Si2}
Z. Ren, L. V. Pourovskii, G. Giriat, G. Lapertot, A. Georges, and D. Jaccard,
Phys. Rev. X {\bf 4}, 031055 (2014).

\bibitem{CeCu2Si2}
F. Steglich, J. Aarts, C. D. Bredl, W. Lieke, D. Meschede, W. Franz, and H. Schafer,
Phys. Rev. Lett. {\bf 43}, 1892 (1979).

\bibitem{spinmediatedpairing}
O. Stockert, J. Arndt, E. Faulhaber, C. Geibel, H. S. Jeevan, S. Kirchner, M. Loewenhaupt, K. Schmalzl, W. Schmidt, Q. Si, and F. Steglich,
Nat. Phys. {\bf 7}, 119 (2011).

\bibitem{valencemediatedpairng}
A. T. Holmes, D. Jaccard, and K. Miyake,
Phys. Rev. B {\bf 69}, 024508 (2004).

\bibitem{Anderson}
P. W. Anderson,
J. Phys. Chem. Solid {\bf 11}, 26 (1959).

\bibitem{UPt3}
Y. Dalichaouch, M. C. de Andrade, D. A. Gajewski, R. Chau, P. Visani, and M. B. Maple,
Phys. Rev. Lett. {\bf 75}, 3938 (1995).

\bibitem{cuprate}
S. K. Tolpygo, J.-Y. Lin, M. Gurvitch, S. Y. Hou, and J. M. Phillips,
Phys. Rev. B {\bf 53}, 12454 (1996).

\bibitem{Sr2RuO4}
A. P. Mackenzie, R. K. W. Haselwimmer, A. W. Tyler, G. G. Lonzarich, Y. Mori, S. Nishizaki, and Y. Maeno,
Phys. Rev. Lett. {\bf 80}, 161 (1998).

\bibitem{organic}
B. J. Powell and R. H. McKenzie,
Phys. Rev. B {\bf 69}, 024519 (2004).

\bibitem{CePt3Si}
M. Nicklas, F. Steglich, J. Knolle, I. Eremin, R. Lackner, and E. Bauer,
Phys. Rev. B {\bf 81}, 180511(R) (2010).

\bibitem{Lonzarich}
P. Monthoux, D. Pines, and G. G. Lonzarich, Nature (London) {\bf 450}, 1177 (2007).

\bibitem{Holmesiron}
A. T. Holmes, D. Jaccard, G. Behr, Y. Inada, and Y. Onuki, J. Phys.:Condens. Matter {\bf 16}, S1121 (2004).

\bibitem{Holmesthesis}
A. T. Holmes, Ph.D. thesis, Universit$\acute{e}$ de Gen$\grave{e}$ve, 2004 [http://archive-ouverte.unige.ch/unige:284].

\bibitem{scattering1}
B. Cornut and B. Coqblin, Phys. Rev. B {\bf 5}, 4541 (1972).

\bibitem{twomaxima}
Y. Nishida, A. Tsuruta, and K. Miyake,
J. Phys. Soc. Jpn. {\bf 75}, 064706 (2006).

\bibitem{determinationofTM}
N. Tateiwa, Y. Haga, T. D. Matsuda, S. Ikeda, T. Yasuda, T. Takeuchi, R. Settai, and Y. Onuki,
J. Phys. Soc. Jpn. {\bf 74}, 1903 (2005).

\bibitem{CeCu2Ge2-1}
D. Jaccard, H. Wilhelm, K. Alami-Yadri, and E. Vargoz,
Physica {\bf 259-261B}, 1 (1999).

\bibitem{CeCu2Si2seyfarth}
G. Seyfarth, A.-S. Ruetschi, K. Sengupta, A. Georges, D. Jaccard, S. Watanabe, and K. Miyake,
Phys. Rev. B {\bf 85}, 205105 (2012).


\bibitem{Holmesreview}
A. T. Holmes, D. Jaccard, and K. Miyake,
J. Phys. Soc. Jpn. {\bf 76}, 051002 (2007).

\bibitem{BellarbiCeCu2Si2}
B. Bellarbi, A. Benoit, D. Jaccard, J. M. Mignot, and H. F. Braun,
Phys. Rev. B {\bf 30}, 1182 (1984).

\bibitem{Miyake}
K. Miyake and O. Narikiyo,
J. Phys. Soc. Jpn. {\bf 71}, 867 (2002).

\bibitem{AG}
A. A. Abrikosov and L. P. Gor'kov,
Sov. Phys. JETP {\bf 12}, 1243 (1961).

\bibitem{AG1}
P. I. Larkin,
Sov. Phys. JETP Lett. {\bf 2}, 130 (1965).

\bibitem{AG2}
A. J. Millis, S. Sachdev, and C. M. Varma,
Phys. Rev. B {\bf 37}, 4975 (1988).

\bibitem{AG3}
R. J. Radtke, K. Levin, H.-B. Sch\"{u}ttler, and M. R. Norman,
Phys. Rev. B {\bf 48}, 653(R) (1993).

\bibitem{UBe13}
M. B. Maple, J. W. Chen, S. E. Lambert, Z. Fisk, J. L. Smith, H. R. Ott, J. S. Brooks, and M. J. Naughton,
Phys. Rev. Lett. {\bf 54}, 477 (1985).

\bibitem{CeCuSiGedisorder}
H. Q. Yuan, F. M. Grosche, M. Deppe, C. Geibel, G. Sparn, and F. Seglich,
New J. Phys. {\bf 6}, 132 (2004).

\bibitem{Emmathesis}
E. Vargoz, Ph.D. thesis, Universit$\acute{e}$ de Gen$\grave{e}$ve, 1998.

\bibitem{CeCu2Si2disorder}
A. T. Holmes, D. Jaccard, H. S. Jeevan, C. Geibel, and M. Ishikawa,
J. Phys.:Condens. Matter {\bf 17}, 5423 (2005).

\bibitem{Kondohole1}
J. M. Lawrence, J. D. Thompson, and Y. Y. Chen,
Phys. Rev. Lett. {\bf 54}, 2537 (1985).

\bibitem{Kondohole2}
D. Jaccard, E. Vargoz, K. Alami-Yadri, and H. Wilhelm,
Rev. High Pressure Sci. Technol. {\bf 7}, 412 (1998).

\bibitem{localorder}
A. J. Millis, D. K. Morr, and J. Schmalian,
Phys. Rev. Lett. {\bf 87}, 167202 (2001).

\bibitem{CdinCeCoIn5}
S. Seo, X. Lu, J.-X. Zhu, R. R. Urbano, N. Curro, E. D. Bauer, V. A. Sidorov, L. D. Pham, T. Park, Z. Fisk, and J. D. Thompson,
Nat. Phys. {\bf 10}, 120 (2014).

\end{thebibliography}
\end{document}